\documentclass[a4paper,fleqn,usenatbib]{mnras}
\usepackage[usenames,dvipsnames]{xcolor}

\usepackage{newtxtext,newtxmath}
\usepackage{tablefootnote}
\usepackage{color}

\usepackage[T1]{fontenc}
\usepackage{ae,aecompl}

\usepackage{graphicx}	
\usepackage{amsmath}	
\usepackage{amssymb}	

\title[Spin Parity of Spiral Galaxies II]{Spin Parity of Spiral Galaxies II: A catalogue of 80k spiral galaxies using big data from the Subaru Hyper Suprime-Cam Survey and deep learning}

\author[K. Tadaki et al.]{
Ken-ichi Tadaki,$^{1}$\thanks{E-mail: tadaki.ken@nao.ac.jp}
Masanori Iye,$^{1}$
Hideya Fukumoto,$^{2}$
Masao Hayashi,$^{1}$
\newauthor
Cristian E. Rusu,$^{1}$
Rhythm Shimakawa,$^{1}$
and Tomoka Tosaki,$^{3}$\\
$^{1}$National Astronomical Observatory of Japan, 2-21-1 Osawa, Mitaka, Tokyo 181-8588, Japan\\
$^{2}$The Open University of Japan, 2-11 Wakaba, Mihama-ku, Chiba 261- 8586 Japan\\
$^{3}$Joetsu University of Education, Yamayashiki-machi, Joetsu, Niigata 943-8512, Japan\\
}

\date{Accepted XXX. Received YYY; in original form ZZZ}

\pubyear{2020}

\begin{document}
\label{firstpage}
\pagerange{\pageref{firstpage}--\pageref{lastpage}}
\maketitle

\begin{abstract}
We report an automated morphological classification of galaxies into S-wise spirals, Z-wise spirals, and non-spirals using big image data taken from Subaru/Hyper Suprime-Cam (HSC) Survey and a convolutional neural network(CNN)-based deep learning technique.
The HSC $i$-band images are about 25 times deeper than those from the Sloan Digital Sky Survey (SDSS) and have a two times higher spatial resolution, allowing us to identify substructures such as spiral arms and bars in galaxies at $z>0.1$.
We train CNN classifiers by using HSC images of 1447 S-spirals, 1382 Z-spirals, and 51,650 non-spirals. 
As the number of images in each class is unbalanced, we augment the data of spiral galaxies by horizontal flipping, rotation, and rescaling of images to make the numbers of three classes similar.
The trained CNN models correctly classify 97.5\% of the validation data, which is not used for training.
We apply the CNNs to HSC images of a half million galaxies with an i-band magnitude of $i<20$ over an area of 320 deg$^2$.
37,917 S-spirals and 38,718 Z-spirals are identified, indicating no significant difference between the numbers of two classes.
Among a total of 76,635 spiral galaxies, 48,576 are located at $z>0.2$, where we are hardly able to identify spiral arms in the SDSS images. 
Our attempt demonstrates that a combination of the HSC big data and CNNs has a large potential to classify various types of morphology such as bars, mergers and strongly-lensed objects.
\end{abstract}




\section{Introduction}

Spiral arms are one of the most beautiful structures in galaxies and attract the interest of many people.
Density wave theory is a traditionally accepted concept to explain spiral patterns in galaxies \citep{1964ApJ...140..646L} whereas recent numerical simulations support that spiral arms are formed due to a swing amplification associated with galactic shear motion \citep[e.g.,][]{2013ApJ...763...46B, 2014PASA...31...35D}, which was originally suggest by \cite{1965MNRAS.130..125G}.
The winding direction of spiral arms with respect to the galaxy rotation direction has also been a classic subject of controversial debates in observational and theoretical studies of spiral galaxies until around the decade of the 1960s.   
\cite{2019ApJ...886..133I} demonstrate a corroborative evidence that all galaxies are trailing spirals provided that the dark lane dominant side is the side of the disk near to us.
Once the winding direction of spiral arms is identified, we can infer the spin vector of galaxies, which is one of important physical properties in the process of galaxy formation.
In the framework of the tidal torque theory \citep[e.g.,][]{1970Ap......6..320D,1969ApJ...155..393P,1984ApJ...286...38W}, galaxies acquire angular momentum by the tidal fields of their neighbors in the linear stage of structure formation. 
Thus, the winding direction, that is the spin vectors, of galaxies would have been randomly located, leading to the isotropic distribution that the number of S-wise spirals (S-spirals) is identical to that of Z-wise spirals (Z-spirals).
If there is a global anisotropy in the spatial distribution of the winding direction, a large-scale vorticity such as galaxy-cluster tidal interaction would affect the spin of galaxies \citep{1995MNRAS.276..327S}.
However, a statistical analysis of the winding direction was not well developed in the past two decades, except for a few studies \citep{2017MNRAS.466.3928H,2017PASA...34...44S}.

We can immediately identify spiral arms in a galaxy and judge whether it is a S-spiral or a Z-spiral by visual inspection.
However, it is not easy to repeat this procedure 10,000 times or more.
Another problem is that visual classification depends on the expertise and experience of people who look at images.
In the Galaxy Zoo project \citep{2011MNRAS.410..166L,2013MNRAS.435.2835W}, about 100,000 volunteers classified the morphology of $\sim$900,000 galaxies at $0.001<z<0.25$, drawn from the Sloan Digital Sky Survey (SDSS; \citealt{2000AJ....120.1579Y}).
\cite{2010MNRAS.405..783M} studied 5,433 face-on spiral galaxies at $0.03<z<0.085$ from the Galaxy Zoo database.
To make a similarly large sample of more distant spiral galaxies at $z>0.1$, we require higher sensitivity and higher resolution imaging data set over a wide area.

We are conducting a multi-band imaging survey by using Hyper Suprime-Cam (HSC) in Subaru Strategic Program (HSC-SSP; \citealt{2018PASJ...70S...4A}).
The HSC has the largest field of view of 1.5 degree diameter on 8-m class telescopes.
The Wide layer of the survey covers 1400 deg$^2$ in five broad bands ($grizy$) with a 5$\sigma$ point-source depth of $i\sim26.2$, which is about 3.5 magnitudes deeper than SDSS. 
The increased sensitivity allows us to characterize spiral arms in distant galaxies at $z>0.1$ while at the same time the wide survey produces images of more than one million galaxies.
We therefore need to develop an automated method for morphological classification in the big data era.
Commonly used parametric methods such as S\'{e}rsic model fitting \citep[e.g.,][]{2010AJ....139.2097P} and nonparametric ones such as the concentration (C), asymmetry (A), clumpiness (S) method and the Gini/M$_{20}$ parameters \citep[e.g.,][]{2014ARA&A..52..291C} are not suitable for identifying substructures in galaxies, such as spiral arms, bars, and tidal streams.
Currently, only several studies succeed in automatically extracting spiral structures \citep{2014ApJ...790...87D, 2016ApJS..223...20K,2017MNRAS.472.2263H}.


In 2012, deep learning has been dramatically developed enough to correctly recognize the picture of a cat as a cat with high accuracy of $\sim$84\% \citep{NIPS2012_4824}.
The accuracy of image classification has exceeded human accuracy in 2015 ($\sim$95\%; \citealt{HeZRS16}).
A convolutional neural network (CNN) is now a commonly-used technique for classifying images into multiple categories \citep[e.g.,][]{Fukushima...1980,LeCun...1998, ILSVRC15}.
CNNs convolve images with multiple kernels (filters) to reduce the amount of information and efficiently extract local features in images.
\cite{2015MNRAS.450.1441D} have applied a CNN technique to astronomical images for galaxy morphology classification and successfully reproduced the results from the Galaxy Zoo with the accuracy of 99\% \citep[see also e.g., ][]{2015ApJS..221....8H, 2018MNRAS.476.3661D, 2018MNRAS.477..894A}.
These approaches are mostly supervised learning, which requires a training data set with pre-labelled images.
On the other hand, there are some studies which adopt an unsupervised learning approach for automated morphological classification \citep[e.g.,][]{2018MNRAS.473.1108H, 2019arXiv190910537M}.
Furthermore, \cite{2017MNRAS.467L.110S} generate super resolution images from artificially degraded low-resolution images using a generative adversarial network \citep{NIPS2014_5423} although they caution about application to unknown galaxy population, which is not included in a training data set.
CNNs and other deep learning techniques are becoming increasingly common in Astronomy.

In this paper, we present CNN models to identify spiral arms in galaxies by using the HSC imaging data.
In Section 2, we describe the imaging data taken from the HSC-SSP survey and build a training data set. 
We show the architecture of CNNs and estimate the accuracy of morphological classification by using validation data sets in Section 3. 
We apply the trained CNNs to an unlabeled data set of a half million galaxy images and present a catalogue of 80k spiral galaxies in Section 4.

\section{Data}


\subsection{Subaru Hyper Suprime Cam Data Sample}

This work is based on data from the second public data release (PDR-2) of the HSC-SSP for the Wide layer \citep{2019PASJ...71..114A}.
For morphological classification, we use $i-$band images, which have reached an exposure time of about 20 minutes.
Figure 1 shows the images from the HSC-SSP data and SDSS for two spiral galaxies, demonstrating the superb image quality with which we can identify the spiral winding sense even in distant galaxies at $z>0.1$.

\begin{figure}
\includegraphics[scale=1]{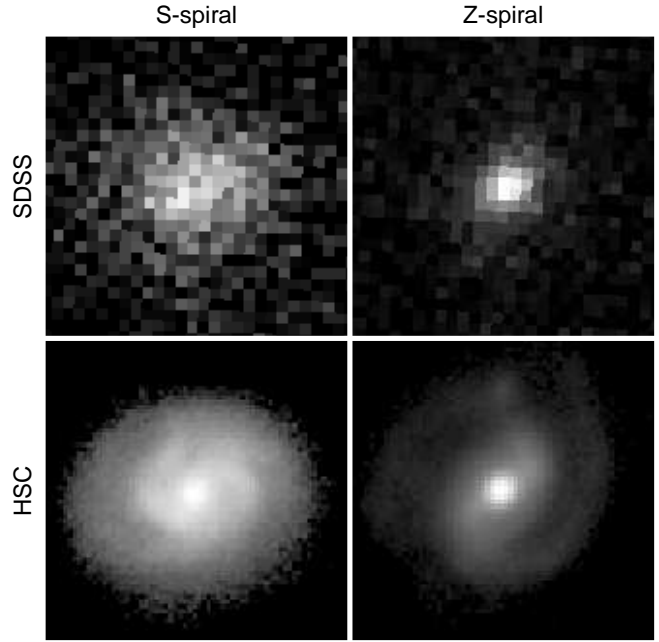}
\caption{Comparisons between SDSS and HSC $i-$band images for a S-wise spiral galaxy (left) with $i=18.7$ at $z=0.16$ and a Z-wise spiral galaxy (right) with $i=18.8$ at $z=0.19$.
The image sizes are all 10.8 arcsec $\times$ 10.8 arcsec.}
\end{figure}

\begin{figure*}
\begin{center}
\includegraphics[scale=1.0]{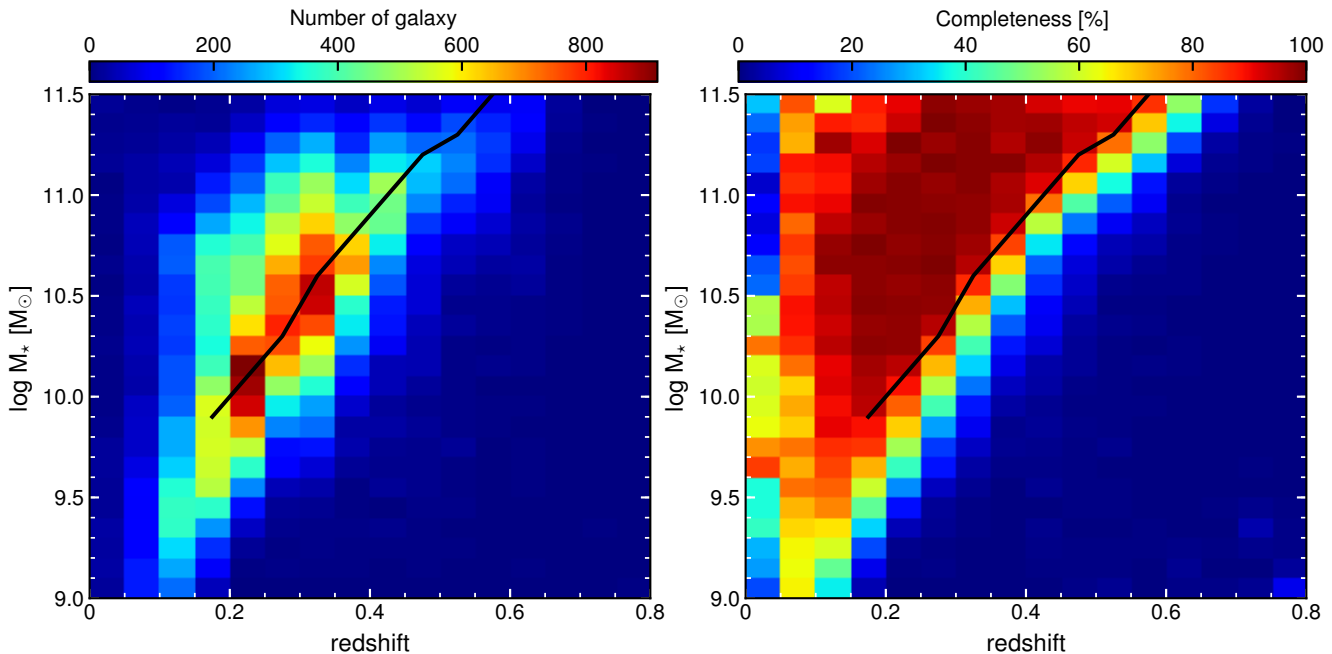}
\end{center}
\caption{Left: stellar mass vs. redshift for galaxies in a subsample of 56,787 galaxies with $i<20$ in the Wide XMM-LSS field.
Right: completeness as functions of stellar mass and redshift.
Black lines denote the completeness of 90\%.
\label{fig;zM}}
\end{figure*}

We use galaxies with 5.7 arcsec aperture magnitudes $i<20$ so that we can visually classify their morphology.
It is still possible to identify spiral arms in HSC images even for galaxies with $i\sim21$, allowing for morphological classification of galaxies at higher redshift.
However, as such case are rare, the inclusion of fainter objects makes it difficult to identify a larger number of spiral galaxies for training CNNs.
We therefore choose a magnitude cut of $i<20$ in this work.

Stars are removed in advance by the flag of {\tt i\_extendedness\_value} in the HSC-SSP database.
As most galaxies are observed in other four broad band filters ($grzy$), their photometric redshift is available, provided by the Direct Empirical Photometric code (DEmP: \citealt{2014ApJ...792..102H}).
For galaxies with $i < 20$, the photometric redshift error of $\Delta z=(z_\mathrm{phot}-z_\mathrm{spec})/(1+z_\mathrm{spec})$ and the outlier fraction of $|\Delta z|>0.15$ is $\sigma_{\Delta z}=0.02$ and 5--10\%, respectively \citep{2018PASJ...70S...9T,2020arXiv200301511N}.
We remove nearby galaxies with spectroscopic redshift of $z_\mathrm{spec}<0.05$, provided by VIPERS \citep{2014A&A...562A..23G}, SDSS \citep{2015ApJS..219...12A}, Wiggle-Z \citep{2010MNRAS.401.1429D}, GAMA \citep{2015MNRAS.452.2087L} and PRIMUS \citep{2013ApJ...767..118C}, as the physical scale resolution becomes significantly different from that at $z>0.1$.
Edge-on like objects with a major-to-minor axis ratio of less than 0.1 are also removed in advance because it is hard to distinguish between spirals and non-spirals.

For a subsample of 56,787 galaxies with $i < 20$ in the Wide XMM Large Scale Structure survey (XMM-LSS) field, the redshift and stellar mass distributions are shown in Figure \ref{fig;zM}.
The stellar mass is estimated from multi-wavelength photometry, empirically given by the DEmP code \citep{2014ApJ...792..102H}. 
We also derive the completeness as functions of redshift and stellar mass by calculating the ratio of the number of galaxies with $i<20$ to the number of fainter galaxies with $i<22$. 
The stellar mass 90\% completeness limits are $\log(M_\star/M_\odot)\sim10$ at $z=0.2$ and $\log(M_\star/M_\odot)\sim11$ at $z=0.4$.

We convert FITS images of galaxies to Joint Photographic Experts Group (JPEG) format by using {\tt STIFF} software \citep{2012ASPC..461..263B}.
We adopt {\tt GAMMA}=2.2 to automatically adjust the contrast and brightness of JPEG images for classification and slightly change this parameter for data augmentation (Section \ref{sec;training}).
Converting images to JPEG images could potentially loose information respect to the original FITS images.
Optimization of the grayscale images is one of the key challenges for improving classification with deep learning, but is beyond the scope of our work.
In this paper, we simply use JPEG images by following the previous works \citep[e.g.,][]{2015MNRAS.450.1441D,2015ApJS..221....8H, 2018MNRAS.476.3661D, 2018MNRAS.477..894A}. 
The size of post stamp images is 64 pixel $\times$ 64 pixel, covering 10.8 arcsec $\times$ 10.8 arcsec, where main spiral features are covered for most galaxies.

Although color composite images are often used for classifying galaxy morphology, we use monochromatic images in the $i$-band for two reasons.
First, $i$-band observations are executed in the best observing conditions for cosmic shear measurements \citep{2018PASJ...70S..25M}. 
The median seeing is 0.6 arcsec in the $i$-band, corresponding to 1.1 kpc at $z=0.1$ and 2.1 kpc at $z=0.2$, while it is 0.7-0.8 arcsec in other bands.
Second, composite images have the information of galaxy colors as well as galaxy morphology.
There is a strong correlation between color and morphology: blue galaxies tend to have a disk with spiral arms while red galaxies are ellipticals \citep[e.g.,][]{2001AJ....122.1861S}.
Red spiral galaxies are likely to be an important population for understanding transitions from blue to red galaxies \citep{2010MNRAS.405..783M} but the number density is smaller compared to blue spiral galaxies.
If color information is taken into account, the trained models would tend to classify red galaxies into non-spirals rather than spirals.
We therefore use $i$-band images to avoid the color bias and keep morphology information independent from colors.

\subsection{Training data set}
\label{sec;training}


\begin{figure}
\includegraphics[scale=1]{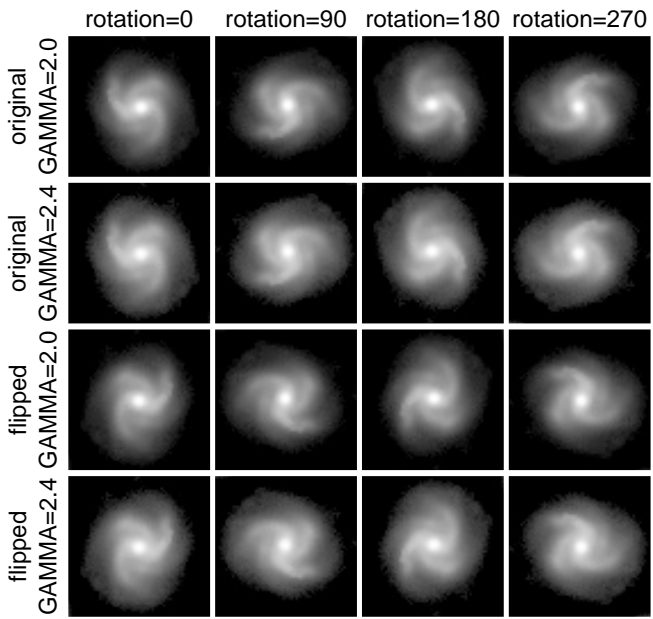}
\caption{Examples of data augmentation. 
A FITS image of one galaxy is converted into multiple JPEG images with a different gray scale value, which is adjusted by the {\tt GAMMA} parameter. 
The original images of S-spirals are horizontally flipped and are treated as Z-spiral images.
Furthermore, these images are rotated by 90, 180, 270 degrees.
\label{fig;dataaug}}
\end{figure}

We chose the XMM-LSS field \citep{2004JCAP...09..011P} over an area of $\sim28$ deg$^2$ to make a training data set for CNNs.
Among 56,787 objects, we confirm 1,447 spiral galaxies with clear S-wise spiral structure and 1,382 with clear Z-wise spiral structure by visual inspection.  
They are visually checked by all of the authors.
Additional 1,177 and 1,131 galaxies are identified to have S-wise and Z-wise spiral structure with somewhat reduced confidence level.
To define as clearly as possible spirals, we classify these galaxies into a category of unclear/dubious, which is not used for training CNNs.
The remaining 51,650 galaxies are non-spiral galaxies.
In this work, we do not distinguish between mergers and non-mergers.
Even if spiral galaxies are clearly affected by tidal interactions with their companions, they are categorized as S-spirals or Z-spirals.
We show example images of randomly-selected 100 galaxies in each class from the training data set in Appendix \ref{sec;SZimages}.

We adopt a K-fold cross validation technique to evaluate the performance of CNN models.
We randomly divide the original sample in each class into five subsets, which consist of 289 S-spirals, 276 Z-spirals, and 10,330 non-spirals.
Four of them (1,156 S-spirals, 1,104 Z-spirals, and 41,320 non-spirals) are used for the training and the remaining one is used for validation.
As the numbers of S-spirals and Z-spirals are much smaller than that of non-spirals, we augment the data of spiral galaxies.
We add horizontally flipped images of Z-spirals and S-spirals to the S-spiral and Z-spiral classes, resulting in the same number in S-spirals and Z-spirals. 
Flipping spiral galaxies is also important for making an unbiased training dataset.
When CNNs are trained by a sample biased to S-spirals, the trained model would naturally give more S-spirals than Z-spirals.
We furthermore rotate the images by 90 degrees, 180 degrees and 270 degrees, and rescale the brightness of the images with {\tt GAMMA}=2.0, 2.1, 2.3, 2.4 parameters (Figure \ref{fig;dataaug}).
The data augmentation increases the number of spiral galaxies by 40 times and make the numbers of three classes similar.
The training data set therefore contains images of 45,200 S-spirals, 45,200 Z-spirals and 41,320 non-spirals.
We eventually make five different training data sets by selecting a different subset for validation.

\section{Convolutional neural networks}

We make CNN models to classify galaxy morphology into non-spirals, S-spirals and Z-spirals in a similar way to previous works \citep[e.g.,][]{2015MNRAS.450.1441D}.
Table \ref{tab;cnn} summarizes the configuration of the CNN used in this paper.
The size of input images is 64 pixel $\times$ 64 pixel.
There are four convolutional layers with kernel sizes of 5 pixel $\times$ 5 pixel, $5\times5$, $3\times3$ and $3\times3$, respectively.
The number of convolutional filters is 32, 64, 128, 128, respectively.
Each filter generates a feature map.
We add two pooling layers, which take the maximum value in 4 pixel $\times$ 4 pixel and $2\times2$.
The maximum pooling efficiently extracts important features like edges as well as reduces the amount of information by resampling.
After convolutional layers, 128 feature maps with 5 pixel $\times$ 5 pixel are flatten and fed into a fully-connected layer with 3200 features.
These features are combined in dense layers.
We also include three dropout layers to avoid overfitting of the CNNs \citep{JMLR:v15:srivastava14a}.
In these layers, 20\%, 50\%, 50\% of input units are randomly set to zero.
The final layer uses the Softmax function, which is computed as

\begin{equation}
  p_{i,c}=e^{s_{i,c}}/\sum_{c=1}^3e^{s_{i,c}},
\end{equation}

\noindent
where $s_{i,c}$ is the output score for the $c$-th category for morphology classification (non-spiral, S-spiral and Z-spiral) of the $i$-th image and $p_{i,c}$ corresponds to the predicted probabilities of each class.
We eventually adopt the class with the highest probability to determine the morphology.

\begin{table}
\caption{Structure of CNNs used in this paper \label{tab;cnn}}
\begin{center}
\begin{tabular}{lcc}
\hline
& layer & output shape \\
\hline
\hline
1 & Input       &   (64 pix, 64 pix, 1 map)      \\
2 & Convolution (5 pix $\times$ 5 pix)  & (60 pix, 60 pix, 32 maps)    \\
3 & Convolution (5 pix $\times$ 5 pix)  &  (56 pix, 56 pix, 64 maps)    \\
4 & MaxPooling (4 pix $\times$ 4 pix)  &   (14 pix, 14 pix, 64 maps)     \\
5 & Convolution (3 pix $\times$ 3 pix)  & (12 pix, 12 pix, 128 maps)    \\
6 & Convolution (3 pix $\times$ 3 pix)  &  (10 pix, 10 pix, 128 maps)   \\
7 & MaxPooling (2 pix $\times$ 2 pix)  & (5 pix, 5 pix, 128 maps)     \\
8 & Dropout  &   (5 pix, 5 pix, 128 maps)   \\
9 & Flatten    &    (3200 features)          \\
10 & Dense    & (3200 features)         \\
11 & Dropout  &   (3200 features)    \\
12 & Dense    &  (3200 features)           \\
13 & Dropout  &   (3200 features)    \\
14 & Dense    &   (3 classes)           \\
\hline
\hline
\end{tabular}
\end{center}
\end{table}

\begin{figure*}
\begin{center}
\includegraphics[scale=0.99]{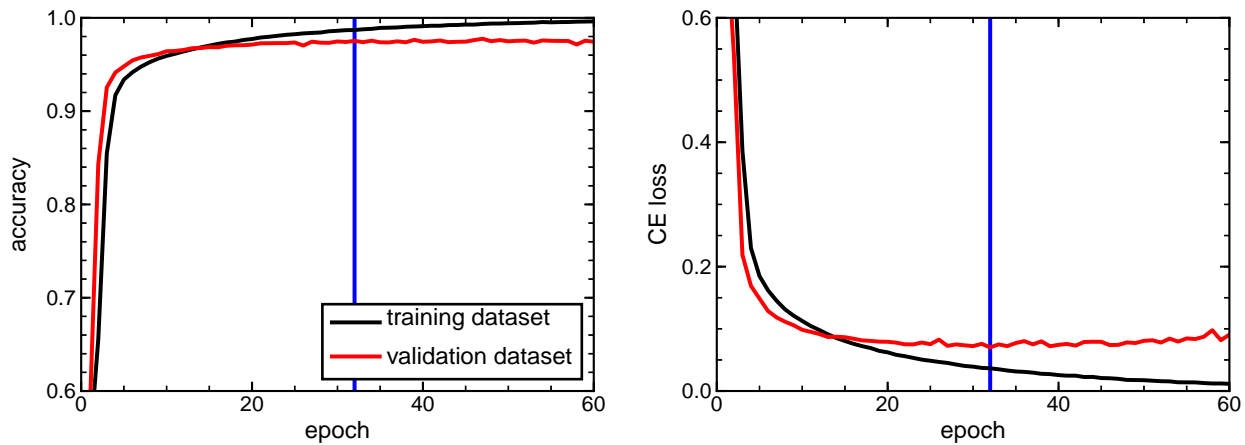}
\end{center}
\caption{The accuracy and the CE loss function for the training (black lines) and validation data set (red lines) for one of the six cross-validation tests.
The blue horizontal line indicate the epoch when the validation loss is minimized.
\label{fig;training}}
\end{figure*}

We train the CNNs using the {\tt Keras} library \citep{chollet2015keras} with a single GPU, NVIDIA GeForce GTX 1080 Ti.
A total of 20,769,539 trainable parameters of the model are determined by minimizing a loss function, which expresses inconsistency between actual classes and predicted probabilities \citep{Krizhevsky14oneweird}. 
We adopt a cross-entropy (CE) loss function defined as,

\begin{equation}
  \mathrm{CE\ loss}=-\sum_{i=1}^{256} \sum_{c=1}^3 t_{i,c}\log p_{i,c},
\end{equation}

\noindent
where $t_{i,c}$ is the ground truth label (1 if true and 0 if false).
We use {\it Adam} algorithm \citep{kingma2014adam}, which optimizes the parameters based on the gradient descent of the loss function with a subsample of 256 images randomly-selected from the training data set.
This method is called the mini-batch stochastic gradient descent.
One epoch ends with 33 seconds when the entire training data set has been used once for the calculation of the loss function.
We repeat this 60 times and derive the accuracy, which is simply a ratio of correctly predicted images to all the images, and the CE loss in each epoch.
The results of one CNN model are shown in Figure \ref{fig;training}.
The training accuracy continues to increase to almost 100\% while the validation accuracy saturates at about 20 epochs.
The loss function takes the minimum at 32 epochs and turns to increase at the later epochs.
As this is clearly overfitting the training data set, we adopt the CNN models where the validation loss is minimized.
Other four CNN models reach the minimum at 29, 32, 41, and 28 epochs.
The validation data set is not directly used for the training of CNNs, but indirectly affects the choice of the best model.
We therefore compute the average accuracy of 5 CNN models from the cross-validation (see section \ref{sec;training}).

\begin{table}
\caption{The fractions of images with the predicted class in each labeled class from the cross-validation. \label{tab;recall}}
\begin{center}
\begin{tabular}{lccc}
\hline
Predicted class & \multicolumn{3}{c}{Labeled class}\\
& non-spiral & S-spiral & Z-spiral  \\
\hline
non-spiral & 96.31$\pm$0.47\%& 1.86$\pm$0.20\% & 1.83$\pm$0.34\%  \\
S-spiral & 1.71$\pm$0.17\% & 98.12$\pm$0.24\% & 0.16$\pm$0.10\% \\
Z-spiral & 1.90$\pm$0.22\% & 0.19$\pm$0.13\% & 97.91$\pm$0.34\% \\
\hline
\end{tabular}
\end{center}
\end{table}

\begin{figure*}
\begin{center}
\includegraphics[scale=1.0]{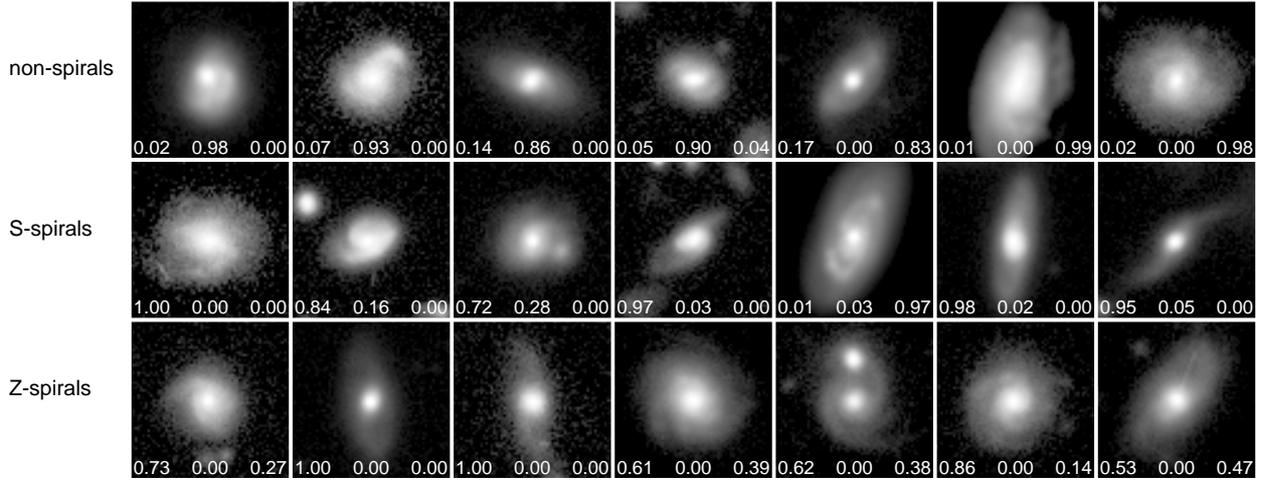}
\end{center}
\caption{Examples of HSC images of misclassification in each class.
From left to right in the bottom of each images, we show the predicted probabilities of non-spiral, S-spiral and Z-spiral.
\label{fig;exam_bad}}
\end{figure*}

\begin{figure*}
\begin{center}
\includegraphics[scale=1.0]{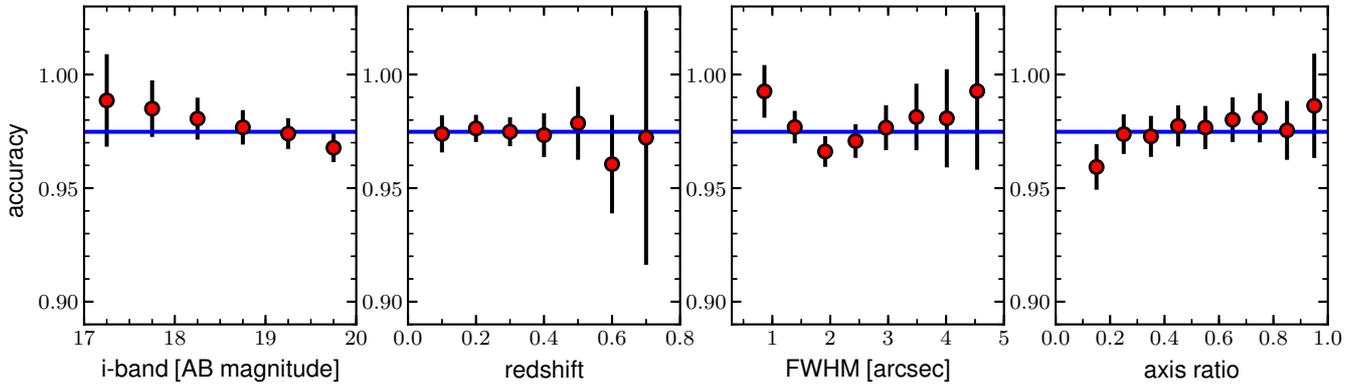}
\end{center}
\caption{The accuracy of the CNN models for the validation data set as a function of $i$-band magnitude, photometric redshift, FWHM size and major-to-minor axis ratio.
A blue line shows the overall accuracy, 0.9748.
\label{fig;acc_val}}
\end{figure*}

The average accuracy and standard deviation is 97.48$\pm$0.14\%.
For galaxy images with the predicted probability of $>$0.95, the accuracy is increased to 99.37$\pm$0.10\%.
Note that the accuracy is different among the morphology classes.
We show the confusion matrix, which is the fraction of correct or incorrect predictions in each predicted class, in Table \ref{tab;recall}.
Most of the failures are for the case that non-spirals are misclassified as either S-spirals or Z-spirals and vice versa.
The fraction that S-spirals (Z-spirals) are misclassified as Z-spirals (S-spirals) is only 0.2\%.

For about 2.5\% of the validation data set, the predicted class is different from the labeled one.
Figure \ref{fig;exam_bad} shows examples of misclassification.
Some objects have the second highest probability of 0.1--0.5 in the labeled class while others are misclassified with the high probability of $>0.95$.
Non-spirals with different predictions seem to have some substructures, suggesting that they potentially have spiral arms with low contrast.
Edge-on galaxies seem to be often misclassified, compared to face-on ones.
The accuracy in the validation data set depends on the major-to-minor axis, which can be interpreted as an inclination angle (Figure \ref{fig;acc_val}).
For galaxies with an axis ratio of $<0.2$, the accuracy decreases to 96\%.
This may be due to a lack of edge-on objects in the training data set as it becomes more difficult to identify spiral arms by visual inspection.

We evaluate the accuracy as a function of $i-$band magnitude, photometric redshift and FWHM size (Figure \ref{fig;acc_val}).
There is also a weak trend that brighter objects are more correctly classified.
On the other hand, the accuracy tends to be constant across a redshift range to $z\sim0.5$.
The FWHM size is estimated from Gaussian-weighted 2nd-order moment in the $i$-band images \citep{2002AJ....123..583B}.
The moment is stored as {\tt i\_sdssshape\_shape11,22,12} in the PDR2 database.  
We compute the determinant radius as $r_\mathrm{det}= (${\tt shape11} $\times$ {\tt shape22} - {\tt shape12}$^2)^{0.25}$.  
Under the assumption of Gaussian, we convert the radius to FWHM by applying $2\sqrt{2\ln 2}$.
The high accuracy in compact galaxies with FWHM$\sim$1\arcsec is due to the fact that most galaxies are classified as non-spirals in the HSC images.

We also look at how the sample size affects the performance of CNNs.
We train CNNs by using images of randomly-selected 100, 200, 400 and 800 galaxies in each class and measure the accuracy of validation data.
When only original images are used for training, the accuracy gradually increases from 52\% at 100 to 90\% at 800.
The data augmentation including horizontal flipping, rotation, and rescaling significantly improves the accuracy from 52\% to 89\% with the same training data set of 100 images.
It requires at least 100 images to reach an accuracy of more than 90\% with the data augmentation.

\section{A catalogue of spiral galaxies}

\begin{table}
\caption{A spin parity Catalogue of spiral galaxies. Object ID is the same in the PDR-2 of the HSC-SSP for the Wide layer \citep{2019PASJ...71..114A}. $p_0$, $p_1$, and $p_2$ indicate the predicted probabilities of non-spirals, S-spirals and Z-spirals, respectively. The full table is available online. \label{tab;Catalogue}}
\begin{center}
\begin{tabular}{lcccc}
\hline
object ID & class flag$^a$ & $p_0$ & $p_1$ & $p_2$ \\
\hline
40959011452899104 & 2 & 0.091 & 0.000 & 0.909 \\ 
40959011452899880 & 2 & 0.093 & 0.000 & 0.907 \\ 
40959011452901552 & 1 & 0.272 & 0.728 & 0.000 \\ 
40959015747870352 & 2 & 0.009 & 0.000 & 0.991 \\ 
40959015747871064 & 1 & 0.011 & 0.989 & 0.000 \\ 
40959020042835800 & 2 & 0.006 & 0.000 & 0.994 \\ 
40959020042837000 & 1 & 0.000 & 1.000 & 0.000 \\ 
. & . & . & . & . \\
\hline
\end{tabular}
\end{center}
$^a$Flag: 1=S-spiral; 2=Z-spiral
\end{table}

\begin{figure*}
\begin{center}
\includegraphics[scale=1.0]{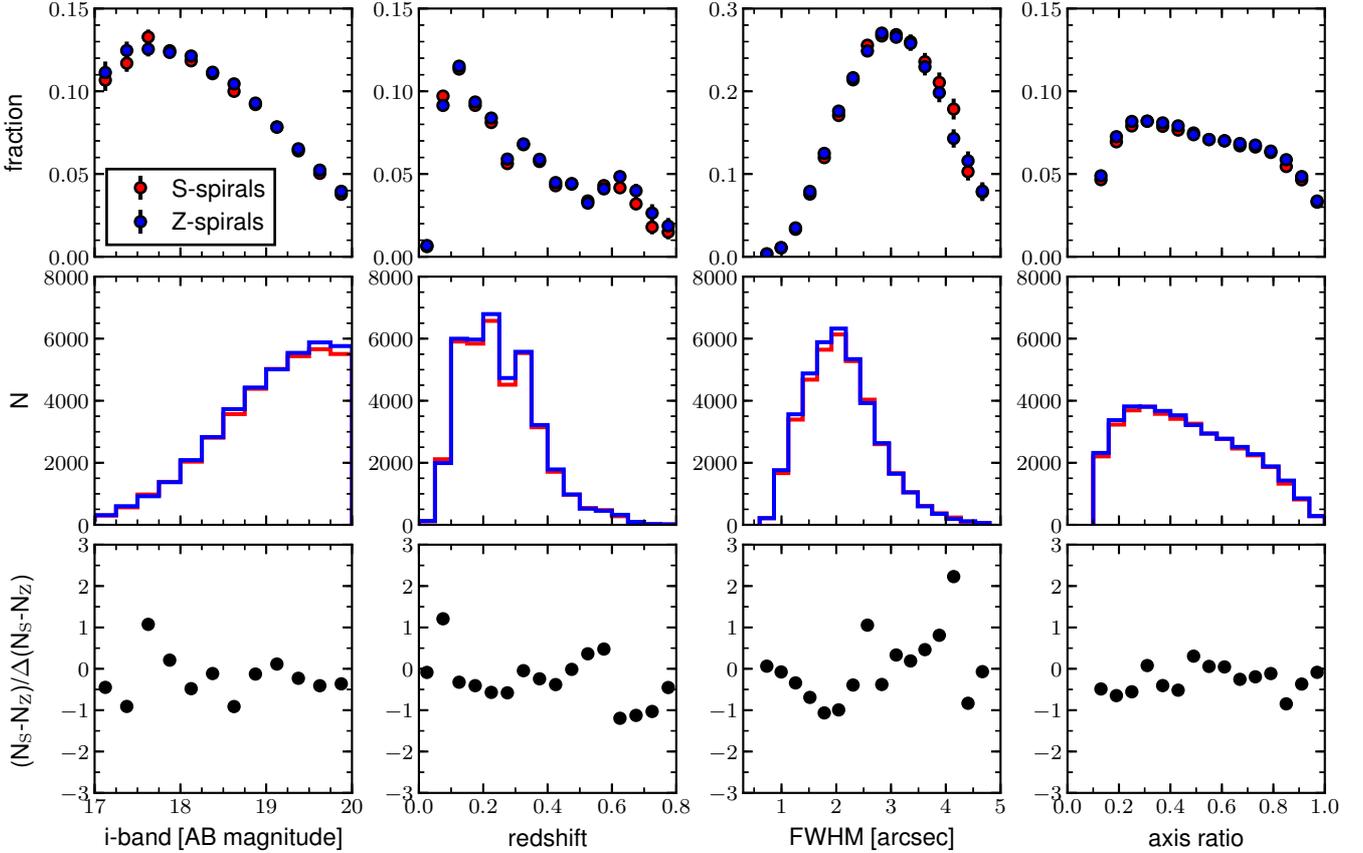}
\end{center}
\caption{
The fraction of S-spirals (red circles) and Z-spirals (blue circles) to all the galaxies as functions of $i$-band magnitude, photometric redshift, FWHM size and major-to-minor axis ratio (top four panels).
The middle four panels and the bottom four panels show the number of spiral galaxies and the significance level of the number difference between S-spirals and Z-spirals, respectively.
The error, $\Delta (N_\mathrm{S}-N_\mathrm{Z})$, takes into account both the Poisson error and the incompleteness of CNN-based classification.
\label{fig;histogram}}
\end{figure*}

Now, we apply the trained CNN models to a large data set in other HSC-SSP fields, where 561,251 galaxy images are available over an area of $\sim$320 deg$^2$.
We use 5 CNN models made from the cross validation to derive average predicted probabilities of each class.
We identify 37,917 S-spirals and 38,718 Z-spirals and provide the catalogue including the predicted probabilities in Table \ref{tab;Catalogue}.
The remaining 484,616 galaxies are non-spirals.
The difference between the numbers of S-spirals and Z-spirals is $N_\mathrm{S}-N_\mathrm{Z}=-801$.
The significance level of is 2.9$\sigma$ when only Poisson statistics is applied to estimate the uncertainties in the number of galaxy images \citep{1986ApJ...303..336G}.
However, the uncertainties on these numbers are likely to be dominated by misclassification of CNN-based classification, rather than Poisson errors.
S-spirals and Z-spirals are in principle affected to the same degree by the contamination.
In the validation data set, the fraction that S-spirals are misclassified as non-spirals is similar to the fraction that Z-spirals are misclassified as non-spirals while there are some variations (0.20\%, 0.34\%) between the CNN models (Table \ref{tab;recall}).
This is non-negligible because the vast majority of a half million galaxies is non-spirals.
Considering the uncertainties in the misclassification, the error of the difference would be $\Delta (N_\mathrm{S}-N_\mathrm{Z})$=1,932, which is larger than the actual measurement.
We also use 5 individual trained CNNs for classification of 561,251 images to calculate the average and the standard deviation of the numbers of spirals, $N_\mathrm{S}=38625\pm1138$ and $N_\mathrm{Z}=39537\pm1479$, corresponding to the error of $\Delta (N_\mathrm{S}-N_\mathrm{Z})$=1866).
A stable performance of 0.04\% in misclassification is required so that the uncertainty is dominated by Poisson errors in the HSC imaging data.
We would need to train the model with more validation data.

We do not find a significant difference of the numbers between S-spirals and Z-spirals.
On the other hand, there is a significant excess of S-spirals over Z-spirals in our training data set and in the Galaxy Zoo Catalogue \citep{2008MNRAS.388.1686L,2008MNRAS.389.1179L}.
This is likely to be caused by a human selection bias \citep{2017MNRAS.466.3928H}.
Visual inspection by human eyes may unconsciously select more S-spirals.
We also calculate the significance level of the number difference between S-spirals and Z-spirals, $(N_\mathrm{S}-N_\mathrm{Z})/\Delta(N_\mathrm{S}-N_\mathrm{Z})$, in bins of $i-$band magnitude, photometric redshift, FWHM size and major-to-minor axis ratio (Figure \ref{fig;histogram}).
We take into account the incompleteness of CNN-based classification (Table \ref{tab;recall}) as well as the Poisson errors.
We do not find a significant excess of S-spirals or Z-spirals with $|(N_\mathrm{S}-N_\mathrm{Z})|>3\Delta(N_\mathrm{S}-N_\mathrm{Z})$.

The fraction of both spirals including S-spirals and Z-spirals to all the galaxies is 13.7\% though it depends on galaxy properties.
Note that we can identify only galaxies with visible spiral arms, depending on the sensitivity and spatial resolution of images used in the classification.
In the Galaxy Zoo project, the fraction of galaxies with features such as spiral arms is 10\% at $z=0.1$ and $\sim0$\% at $z=0.2$ \citep{2013MNRAS.435.2835W}.
In the deeper HSC images, spiral arms are visible in $\sim$20\% of galaxies at $z=0.1-0.2$ (Figure \ref{fig;histogram}).
The measured fraction of spiral galaxies should be still a lower limit because more galaxies with fainter or lower contrast spiral arms can be identified in even deeper and higher-resolution images.
We actually find that $\sim$50\% of extended galaxies with FWHM$\sim$3 arcsec are classified as spirals since the HSC resolution is high enough to identify their spiral arms.

\begin{figure*}
\begin{center}
\includegraphics[scale=1.0]{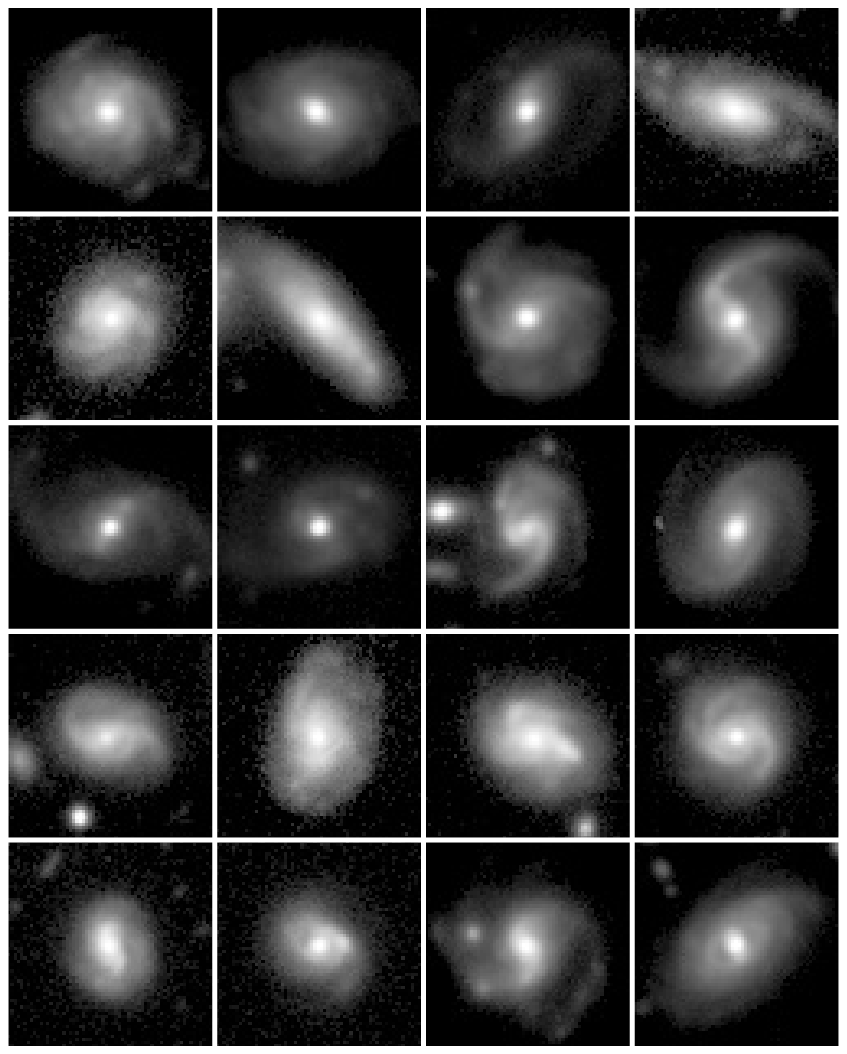}
\includegraphics[scale=1.0]{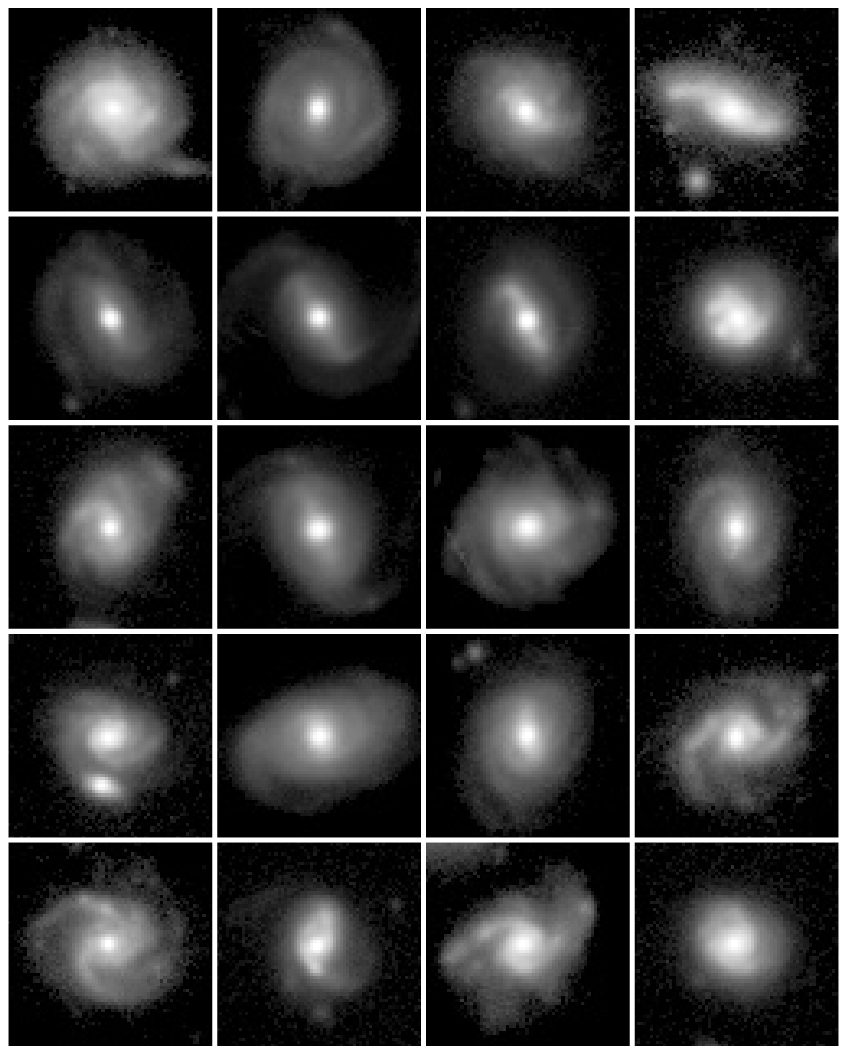}
\end{center}
\caption{Examples of HSC images of S-spirals (left) and Z-spirals (right) with the predicted probability of $>$0.95. They are randomly selected from spectroscopically-confirmed galaxies at $z_\mathrm{spec}=0.2-0.3$.
\label{fig;spiral_z0.2}}
\end{figure*}

Figure \ref{fig;spiral_z0.2} presents examples of S-spirals and Z-spirals with a spectroscopic redshift of $z_\mathrm{spec}=0.2-0.3$.
48,576 of 76,635 spiral galaxies are located at $z_\mathrm{phot}>0.2$, where we are not able to identify spiral arms with the SDSS images.
The fraction of spiral galaxies decreases from 20\% at $z_\mathrm{phot}=0.2$ to 10\% at $z_\mathrm{phot}=0.6$.
The redshift dependence is strongly affected by the cosmological dimming of the surface brightness, which decreases as $(1+z)^{-4}$.
It becomes difficult to detect an extended substructure such as spiral arms in high-redshift galaxies.
The magnitude dependence of the spiral fraction is coupled with the redshift dependence since faint sources tend to be at higher redshift.
The decrease of spirals with a small axis ratio of $<0.2$ is likely to be caused by human bias since it becomes hard to visually identify spiral arms in edge-on galaxies.

2,524 galaxies are identified to have spiral arms at $z_\mathrm{phot}=0.5-0.7$ in spite of the strong effect of the cosmological dimming.
1,455 of them have a stellar mass of $\log(M_\star/M_\odot)>10.8$, which is similar to that of Andromeda (M31: \citealt{2006MNRAS.366..996G}).
The existence of spiral arms indicates that the galaxies are still forming stars.
The majority of M31-mass galaxies have an early type morphology without spiral arms at $0<z<0.7$ and quench the star formation \citep{2015ApJ...803...26P}.
The identified massive spiral galaxies are likely to be the progenitors of M31.

\section{Summary}

We have developed a CNN model to classify galaxy morphology into three categories (S-spiral, Z-spiral, and non-spiral) by using images taken by the Subaru HSC survey.
The superb image quality allows us to identify spiral arms in faint galaxies with $i\sim20$ by visual inspection.
We have used a total of 0.2 million images after data augmentation such as flipping, rescaling, and rotation for training the model.
The trained model successfully classifies the test data set, which is not used for training and validation, and results in an accuracy of 97.5$\pm$0.1\%.
The accuracy decreases to $\sim$90\% in the case that the training data set consists of less than 100 images in each class.
This would become more of a problem when one finds rare objects. 

We have applied the trained CNN model to 561,251 galaxy images over an area of $\sim$320 deg$^2$.
Our automated classification efficiently picks up spiral galaxies and determines their winding direction of spiral arms, providing 37,917 S-spirals and 38,718 Z-spirals.
We do not find a significant excess of S-spirals over Z-spirals, which is seen in the training data set and the Galaxy Zoo Catalogue.
We have also identified 1,455 massive spiral galaxies with $\log(M_\star/M_\odot)>10.8$ at $z_\mathrm{phot}=0.5-0.7$, which are likely the progenitors of M31.
 
There are some limitations to our CNN-based classification.
The criterion of spiral arms is defined by the training data set, which is selected by our visual inspection.
Although we have used the sample of galaxies whose spiral arms are clearly seen for the training to minimize the contamination of non-spirals, the criterion of clear spirals is still somewhat ambiguous.
The ambiguous definition is probably one of the reasons that 2.5\% of the validation data set is misclassified.
It would be important to make a clean training sample.
Creating mock images from numerical simulations is one of several efficient methods to prepare a large data set for training models \citep[e.g.,][]{2018ApJ...858..114H,2019A&A...625A.119M}.
Another direction to define a repeatable class of morphology is an unsupervised learning approach, which does not require visually-classified training data sets \citep[e.g.,][]{2018MNRAS.473.1108H, 2019arXiv190910537M}.
Nevertheless, our attempt already demonstrates that CNN is powerful for making a large sample of galaxies with particular substructures such as spiral arms from a large data-set and efficiently picking up rare objects such as massive spiral galaxies.

\section*{Acknowledgements}

We are very grateful to the referee for constructive suggestions to improve the paper.
The Hyper Suprime-Cam (HSC) collaboration includes the astronomical communities of Japan and Taiwan, and Princeton University.  The HSC instrumentation and software were developed by the National Astronomical Observatory of Japan (NAOJ), the Kavli Institute for the Physics and Mathematics of the Universe (Kavli IPMU), the University of Tokyo, the High Energy Accelerator Research Organization (KEK), the Academia Sinica Institute for Astronomy and Astrophysics in Taiwan (ASIAA), and Princeton University.  Funding was contributed by the FIRST program from the Japanese Cabinet Office, the Ministry of Education, Culture, Sports, Science and Technology (MEXT), the Japan Society for the Promotion of Science (JSPS), Japan Science and Technology Agency  (JST), the Toray Science  Foundation, NAOJ, Kavli IPMU, KEK, ASIAA, and Princeton University.
 
This paper makes use of software developed for the Large Synoptic Survey Telescope. We thank the LSST Project for making their code available as free software at  http://dm.lsst.org
 
This paper is based on data collected at the Subaru Telescope and retrieved from the HSC data archive system, which is operated by Subaru Telescope and Astronomy Data Center (ADC) at NAOJ. 
Data analysis was in part carried out with the cooperation of Center for Computational Astrophysics (CfCA), NAOJ.
 
The Pan-STARRS1 Surveys (PS1) and the PS1 public science archive have been made possible through contributions by the Institute for Astronomy, the University of Hawaii, the Pan-STARRS Project Office, the Max Planck Society and its participating institutes, the Max Planck Institute for Astronomy, Heidelberg, and the Max Planck Institute for Extraterrestrial Physics, Garching, The Johns Hopkins University, Durham University, the University of Edinburgh, the Queen’s University Belfast, the Harvard-Smithsonian Center for Astrophysics, the Las Cumbres Observatory Global Telescope Network Incorporated, the National Central University of Taiwan, the Space Telescope Science Institute, the National Aeronautics and Space Administration under grant No. NNX08AR22G issued through the Planetary Science Division of the NASA Science Mission Directorate, the National Science Foundation grant No. AST-1238877, the University of Maryland, Eotvos Lorand University (ELTE), the Los Alamos National Laboratory, and the Gordon and Betty Moore Foundation.

C.E.R acknowledges Anupreeta More for providing a tool for visual inspection of images.





\begin{thebibliography}{}
\makeatletter
\relax
\def\mn@urlcharsother{\let\do\@makeother \do\$\do\&\do\#\do\^\do\_\do\%\do\~}
\def\mn@doi{\begingroup\mn@urlcharsother \@ifnextchar [ {\mn@doi@}
  {\mn@doi@[]}}
\def\mn@doi@[#1]#2{\def\@tempa{#1}\ifx\@tempa\@empty \href
  {http://dx.doi.org/#2} {doi:#2}\else \href {http://dx.doi.org/#2} {#1}\fi
  \endgroup}
\def\mn@eprint#1#2{\mn@eprint@#1:#2::\@nil}
\def\mn@eprint@arXiv#1{\href {http://arxiv.org/abs/#1} {{\tt arXiv:#1}}}
\def\mn@eprint@dblp#1{\href {http://dblp.uni-trier.de/rec/bibtex/#1.xml}
  {dblp:#1}}
\def\mn@eprint@#1:#2:#3:#4\@nil{\def\@tempa {#1}\def\@tempb {#2}\def\@tempc
  {#3}\ifx \@tempc \@empty \let \@tempc \@tempb \let \@tempb \@tempa \fi \ifx
  \@tempb \@empty \def\@tempb {arXiv}\fi \@ifundefined
  {mn@eprint@\@tempb}{\@tempb:\@tempc}{\expandafter \expandafter \csname
  mn@eprint@\@tempb\endcsname \expandafter{\@tempc}}}

\bibitem[\protect\citeauthoryear{{Abraham}, {Aniyan}, {Kembhavi}, {Philip}  \&
  {Vaghmare}}{{Abraham} et~al.}{2018}]{2018MNRAS.477..894A}
{Abraham} S.,  {Aniyan} A.~K.,  {Kembhavi} A.~K.,  {Philip} N.~S.,   {Vaghmare}
  K.,  2018, \mn@doi [\mnras] {10.1093/mnras/sty627}, \href
  {https://ui.adsabs.harvard.edu/abs/2018MNRAS.477..894A} {477, 894}

\bibitem[\protect\citeauthoryear{{Aihara} et~al.,}{{Aihara}
  et~al.}{2018}]{2018PASJ...70S...4A}
{Aihara} H.,  et~al., 2018, \mn@doi [\pasj] {10.1093/pasj/psx066}, \href
  {https://ui.adsabs.harvard.edu/abs/2018PASJ...70S...4A} {70, S4}

\bibitem[\protect\citeauthoryear{{Aihara} et~al.,}{{Aihara}
  et~al.}{2019}]{2019PASJ...71..114A}
{Aihara} H.,  et~al., 2019, \mn@doi [\pasj] {10.1093/pasj/psz103}, \href
  {https://ui.adsabs.harvard.edu/abs/2019PASJ...71..114A} {71, 114}

\bibitem[\protect\citeauthoryear{{Alam} et~al.,}{{Alam}
  et~al.}{2015}]{2015ApJS..219...12A}
{Alam} S.,  et~al., 2015, \mn@doi [\apjs] {10.1088/0067-0049/219/1/12}, \href
  {https://ui.adsabs.harvard.edu/abs/2015ApJS..219...12A} {219, 12}

\bibitem[\protect\citeauthoryear{{Baba}, {Saitoh}  \& {Wada}}{{Baba}
  et~al.}{2013}]{2013ApJ...763...46B}
{Baba} J.,  {Saitoh} T.~R.,   {Wada} K.,  2013, \mn@doi [\apj]
  {10.1088/0004-637X/763/1/46}, \href
  {https://ui.adsabs.harvard.edu/abs/2013ApJ...763...46B} {763, 46}

\bibitem[\protect\citeauthoryear{{Bernstein} \& {Jarvis}}{{Bernstein} \&
  {Jarvis}}{2002}]{2002AJ....123..583B}
{Bernstein} G.~M.,  {Jarvis} M.,  2002, \mn@doi [\aj] {10.1086/338085}, \href
  {https://ui.adsabs.harvard.edu/abs/2002AJ....123..583B} {123, 583}

\bibitem[\protect\citeauthoryear{{Bertin}}{{Bertin}}{2012}]{2012ASPC..461..263B}
{Bertin} E.,  2012, in {Ballester} P.,  {Egret} D.,   {Lorente} N.~P.~F.,  eds,
   Astronomical Society of the Pacific Conference Series Vol. 461, Astronomical
  Data Analysis Software and Systems XXI. p.~263

\bibitem[\protect\citeauthoryear{Chollet et~al.}{Chollet
  et~al.}{2015}]{chollet2015keras}
Chollet F.,  et~al., 2015, Keras, \url{https://keras.io}

\bibitem[\protect\citeauthoryear{{Conselice}}{{Conselice}}{2014}]{2014ARA&A..52..291C}
{Conselice} C.~J.,  2014, \mn@doi [\araa]
  {10.1146/annurev-astro-081913-040037}, \href
  {https://ui.adsabs.harvard.edu/abs/2014ARA&A..52..291C} {52, 291}

\bibitem[\protect\citeauthoryear{{Cool} et~al.,}{{Cool}
  et~al.}{2013}]{2013ApJ...767..118C}
{Cool} R.~J.,  et~al., 2013, \mn@doi [\apj] {10.1088/0004-637X/767/2/118},
  \href {https://ui.adsabs.harvard.edu/abs/2013ApJ...767..118C} {767, 118}

\bibitem[\protect\citeauthoryear{{Davis} \& {Hayes}}{{Davis} \&
  {Hayes}}{2014}]{2014ApJ...790...87D}
{Davis} D.~R.,  {Hayes} W.~B.,  2014, \mn@doi [\apj]
  {10.1088/0004-637X/790/2/87}, \href
  {https://ui.adsabs.harvard.edu/abs/2014ApJ...790...87D} {790, 87}

\bibitem[\protect\citeauthoryear{{Dieleman}, {Willett}  \& {Dambre}}{{Dieleman}
  et~al.}{2015}]{2015MNRAS.450.1441D}
{Dieleman} S.,  {Willett} K.~W.,   {Dambre} J.,  2015, \mn@doi [\mnras]
  {10.1093/mnras/stv632}, \href
  {https://ui.adsabs.harvard.edu/abs/2015MNRAS.450.1441D} {450, 1441}

\bibitem[\protect\citeauthoryear{{Dobbs} \& {Baba}}{{Dobbs} \&
  {Baba}}{2014}]{2014PASA...31...35D}
{Dobbs} C.,  {Baba} J.,  2014, \mn@doi [\pasa] {10.1017/pasa.2014.31}, \href
  {https://ui.adsabs.harvard.edu/abs/2014PASA...31...35D} {31, e035}

\bibitem[\protect\citeauthoryear{{Dom{\'\i}nguez S{\'a}nchez},
  {Huertas-Company}, {Bernardi}, {Tuccillo}  \& {Fischer}}{{Dom{\'\i}nguez
  S{\'a}nchez} et~al.}{2018}]{2018MNRAS.476.3661D}
{Dom{\'\i}nguez S{\'a}nchez} H.,  {Huertas-Company} M.,  {Bernardi} M.,
  {Tuccillo} D.,   {Fischer} J.~L.,  2018, \mn@doi [\mnras]
  {10.1093/mnras/sty338}, \href
  {https://ui.adsabs.harvard.edu/abs/2018MNRAS.476.3661D} {476, 3661}

\bibitem[\protect\citeauthoryear{{Doroshkevich}}{{Doroshkevich}}{1970}]{1970Ap......6..320D}
{Doroshkevich} A.~G.,  1970, \mn@doi [Astrophysics] {10.1007/BF01001625}, \href
  {https://ui.adsabs.harvard.edu/abs/1970Ap......6..320D} {6, 320}

\bibitem[\protect\citeauthoryear{{Drinkwater} et~al.,}{{Drinkwater}
  et~al.}{2010}]{2010MNRAS.401.1429D}
{Drinkwater} M.~J.,  et~al., 2010, \mn@doi [\mnras]
  {10.1111/j.1365-2966.2009.15754.x}, \href
  {https://ui.adsabs.harvard.edu/abs/2010MNRAS.401.1429D} {401, 1429}

\bibitem[\protect\citeauthoryear{Fukushima}{Fukushima}{1980}]{Fukushima...1980}
Fukushima K.,  1980, Biological Cybernetics, 36, 193

\bibitem[\protect\citeauthoryear{{Garilli} et~al.,}{{Garilli}
  et~al.}{2014}]{2014A&A...562A..23G}
{Garilli} B.,  et~al., 2014, \mn@doi [\aap] {10.1051/0004-6361/201322790},
  \href {https://ui.adsabs.harvard.edu/abs/2014A&A...562A..23G} {562, A23}

\bibitem[\protect\citeauthoryear{{Geehan}, {Fardal}, {Babul}  \&
  {Guhathakurta}}{{Geehan} et~al.}{2006}]{2006MNRAS.366..996G}
{Geehan} J.~J.,  {Fardal} M.~A.,  {Babul} A.,   {Guhathakurta} P.,  2006,
  \mn@doi [\mnras] {10.1111/j.1365-2966.2005.09863.x}, \href
  {https://ui.adsabs.harvard.edu/abs/2006MNRAS.366..996G} {366, 996}

\bibitem[\protect\citeauthoryear{{Gehrels}}{{Gehrels}}{1986}]{1986ApJ...303..336G}
{Gehrels} N.,  1986, \mn@doi [\apj] {10.1086/164079}, \href
  {https://ui.adsabs.harvard.edu/abs/1986ApJ...303..336G} {303, 336}

\bibitem[\protect\citeauthoryear{{Goldreich} \& {Lynden-Bell}}{{Goldreich} \&
  {Lynden-Bell}}{1965}]{1965MNRAS.130..125G}
{Goldreich} P.,  {Lynden-Bell} D.,  1965, \mn@doi [\mnras]
  {10.1093/mnras/130.2.125}, \href
  {https://ui.adsabs.harvard.edu/abs/1965MNRAS.130..125G} {130, 125}

\bibitem[\protect\citeauthoryear{Goodfellow, Pouget-Abadie, Mirza, Xu,
  Warde-Farley, Ozair, Courville  \& Bengio}{Goodfellow
  et~al.}{2014}]{NIPS2014_5423}
Goodfellow I.,  Pouget-Abadie J.,  Mirza M.,  Xu B.,  Warde-Farley D.,  Ozair
  S.,  Courville A.,   Bengio Y.,  2014, in Ghahramani Z.,  Welling M.,  Cortes
  C.,  Lawrence N.~D.,   Weinberger K.~Q.,  eds, , Advances in Neural
  Information Processing Systems 27.
Curran Associates, Inc., pp 2672--2680

\bibitem[\protect\citeauthoryear{{Hart} et~al.,}{{Hart}
  et~al.}{2017}]{2017MNRAS.472.2263H}
{Hart} R.~E.,  et~al., 2017, \mn@doi [\mnras] {10.1093/mnras/stx2137}, \href
  {https://ui.adsabs.harvard.edu/abs/2017MNRAS.472.2263H} {472, 2263}

\bibitem[\protect\citeauthoryear{{Hayes}, {Davis}  \& {Silva}}{{Hayes}
  et~al.}{2017}]{2017MNRAS.466.3928H}
{Hayes} W.~B.,  {Davis} D.,   {Silva} P.,  2017, \mn@doi [\mnras]
  {10.1093/mnras/stw3290}, \href
  {https://ui.adsabs.harvard.edu/abs/2017MNRAS.466.3928H} {466, 3928}

\bibitem[\protect\citeauthoryear{He, Zhang, Ren  \& Sun}{He
  et~al.}{2016}]{HeZRS16}
He K.,  Zhang X.,  Ren S.,   Sun J.,  2016, in 2016 {IEEE} Conference on
  Computer Vision and Pattern Recognition, {CVPR} 2016, Las Vegas, NV, USA,
  June 27-30, 2016. pp 770--778, \mn@doi{10.1109/CVPR.2016.90}

\bibitem[\protect\citeauthoryear{{Hocking}, {Geach}, {Sun}  \&
  {Davey}}{{Hocking} et~al.}{2018}]{2018MNRAS.473.1108H}
{Hocking} A.,  {Geach} J.~E.,  {Sun} Y.,   {Davey} N.,  2018, \mn@doi [\mnras]
  {10.1093/mnras/stx2351}, \href
  {https://ui.adsabs.harvard.edu/abs/2018MNRAS.473.1108H} {473, 1108}

\bibitem[\protect\citeauthoryear{{Hsieh} \& {Yee}}{{Hsieh} \&
  {Yee}}{2014}]{2014ApJ...792..102H}
{Hsieh} B.~C.,  {Yee} H.~K.~C.,  2014, \mn@doi [\apj]
  {10.1088/0004-637X/792/2/102}, \href
  {https://ui.adsabs.harvard.edu/abs/2014ApJ...792..102H} {792, 102}

\bibitem[\protect\citeauthoryear{{Huertas-Company} et~al.,}{{Huertas-Company}
  et~al.}{2015}]{2015ApJS..221....8H}
{Huertas-Company} M.,  et~al., 2015, \mn@doi [\apjs]
  {10.1088/0067-0049/221/1/8}, \href
  {https://ui.adsabs.harvard.edu/abs/2015ApJS..221....8H} {221, 8}

\bibitem[\protect\citeauthoryear{{Huertas-Company} et~al.,}{{Huertas-Company}
  et~al.}{2018}]{2018ApJ...858..114H}
{Huertas-Company} M.,  et~al., 2018, \mn@doi [\apj] {10.3847/1538-4357/aabfed},
  \href {https://ui.adsabs.harvard.edu/abs/2018ApJ...858..114H} {858, 114}

\bibitem[\protect\citeauthoryear{{Iye}, {Tadaki}  \& {Fukumoto}}{{Iye}
  et~al.}{2019}]{2019ApJ...886..133I}
{Iye} M.,  {Tadaki} K.-i.,   {Fukumoto} H.,  2019, \mn@doi [\apj]
  {10.3847/1538-4357/ab4a18}, \href
  {https://ui.adsabs.harvard.edu/abs/2019ApJ...886..133I} {886, 133}

\bibitem[\protect\citeauthoryear{Kingma \& Ba}{Kingma \&
  Ba}{2014}]{kingma2014adam}
Kingma D.~P.,  Ba J.,  2014, arXiv preprint arXiv:1412.6980

\bibitem[\protect\citeauthoryear{Krizhevsky \& Inc}{Krizhevsky \&
  Inc}{2014}]{Krizhevsky14oneweird}
Krizhevsky A.,  Inc G.,  2014, Technical report, One weird trick for
  parallelizing convolutional neural networks

\bibitem[\protect\citeauthoryear{Krizhevsky, Sutskever  \& Hinton}{Krizhevsky
  et~al.}{2012}]{NIPS2012_4824}
Krizhevsky A.,  Sutskever I.,   Hinton G.~E.,  2012, in Pereira F.,  Burges C.
  J.~C.,  Bottou L.,   Weinberger K.~Q.,  eds, , Advances in Neural Information
  Processing Systems 25.
Curran Associates, Inc., pp 1097--1105

\bibitem[\protect\citeauthoryear{{Kuminski} \& {Shamir}}{{Kuminski} \&
  {Shamir}}{2016}]{2016ApJS..223...20K}
{Kuminski} E.,  {Shamir} L.,  2016, \mn@doi [\apjs]
  {10.3847/0067-0049/223/2/20}, \href
  {https://ui.adsabs.harvard.edu/abs/2016ApJS..223...20K} {223, 20}

\bibitem[\protect\citeauthoryear{{Land} et~al.,}{{Land}
  et~al.}{2008}]{2008MNRAS.388.1686L}
{Land} K.,  et~al., 2008, \mn@doi [\mnras] {10.1111/j.1365-2966.2008.13490.x},
  \href {https://ui.adsabs.harvard.edu/abs/2008MNRAS.388.1686L} {388, 1686}

\bibitem[\protect\citeauthoryear{LeCun, Bottou, Bengio  \& Haffner}{LeCun
  et~al.}{1998}]{LeCun...1998}
LeCun Y.,  Bottou L.,  Bengio Y.,   Haffner P.,  1998, \mn@doi [Proceedings of
  the IEEE] {10.1109/5.726791}, 86, 2278

\bibitem[\protect\citeauthoryear{{Lin} \& {Shu}}{{Lin} \&
  {Shu}}{1964}]{1964ApJ...140..646L}
{Lin} C.~C.,  {Shu} F.~H.,  1964, \mn@doi [\apj] {10.1086/147955}, \href
  {https://ui.adsabs.harvard.edu/abs/1964ApJ...140..646L} {140, 646}

\bibitem[\protect\citeauthoryear{{Lintott} et~al.,}{{Lintott}
  et~al.}{2008}]{2008MNRAS.389.1179L}
{Lintott} C.~J.,  et~al., 2008, \mn@doi [\mnras]
  {10.1111/j.1365-2966.2008.13689.x}, \href
  {https://ui.adsabs.harvard.edu/abs/2008MNRAS.389.1179L} {389, 1179}

\bibitem[\protect\citeauthoryear{{Lintott} et~al.,}{{Lintott}
  et~al.}{2011}]{2011MNRAS.410..166L}
{Lintott} C.,  et~al., 2011, \mn@doi [\mnras]
  {10.1111/j.1365-2966.2010.17432.x}, \href
  {https://ui.adsabs.harvard.edu/abs/2011MNRAS.410..166L} {410, 166}

\bibitem[\protect\citeauthoryear{{Liske} et~al.,}{{Liske}
  et~al.}{2015}]{2015MNRAS.452.2087L}
{Liske} J.,  et~al., 2015, \mn@doi [\mnras] {10.1093/mnras/stv1436}, \href
  {https://ui.adsabs.harvard.edu/abs/2015MNRAS.452.2087L} {452, 2087}

\bibitem[\protect\citeauthoryear{{Mandelbaum} et~al.,}{{Mandelbaum}
  et~al.}{2018}]{2018PASJ...70S..25M}
{Mandelbaum} R.,  et~al., 2018, \mn@doi [\pasj] {10.1093/pasj/psx130}, \href
  {https://ui.adsabs.harvard.edu/abs/2018PASJ...70S..25M} {70, S25}

\bibitem[\protect\citeauthoryear{{Martin}, {Kaviraj}, {Hocking}, {Read}  \&
  {Geach}}{{Martin} et~al.}{2019}]{2019arXiv190910537M}
{Martin} G.,  {Kaviraj} S.,  {Hocking} A.,  {Read} S.~C.,   {Geach} J.~E.,
  2019, arXiv e-prints, \href
  {https://ui.adsabs.harvard.edu/abs/2019arXiv190910537M} {p. arXiv:1909.10537}

\bibitem[\protect\citeauthoryear{{Masters} et~al.,}{{Masters}
  et~al.}{2010}]{2010MNRAS.405..783M}
{Masters} K.~L.,  et~al., 2010, \mn@doi [\mnras]
  {10.1111/j.1365-2966.2010.16503.x}, \href
  {https://ui.adsabs.harvard.edu/abs/2010MNRAS.405..783M} {405, 783}

\bibitem[\protect\citeauthoryear{{Metcalf} et~al.,}{{Metcalf}
  et~al.}{2019}]{2019A&A...625A.119M}
{Metcalf} R.~B.,  et~al., 2019, \mn@doi [\aap] {10.1051/0004-6361/201832797},
  \href {https://ui.adsabs.harvard.edu/abs/2019A&A...625A.119M} {625, A119}

\bibitem[\protect\citeauthoryear{{Nishizawa}, {Hsieh}, {Tanaka}  \&
  {Takata}}{{Nishizawa} et~al.}{2020}]{2020arXiv200301511N}
{Nishizawa} A.~J.,  {Hsieh} B.-C.,  {Tanaka} M.,   {Takata} T.,  2020, arXiv
  e-prints, \href {https://ui.adsabs.harvard.edu/abs/2020arXiv200301511N} {p.
  arXiv:2003.01511}

\bibitem[\protect\citeauthoryear{{Papovich} et~al.,}{{Papovich}
  et~al.}{2015}]{2015ApJ...803...26P}
{Papovich} C.,  et~al., 2015, \mn@doi [\apj] {10.1088/0004-637X/803/1/26},
  \href {https://ui.adsabs.harvard.edu/abs/2015ApJ...803...26P} {803, 26}

\bibitem[\protect\citeauthoryear{{Peebles}}{{Peebles}}{1969}]{1969ApJ...155..393P}
{Peebles} P.~J.~E.,  1969, \mn@doi [\apj] {10.1086/149876}, \href
  {https://ui.adsabs.harvard.edu/abs/1969ApJ...155..393P} {155, 393}

\bibitem[\protect\citeauthoryear{{Peng}, {Ho}, {Impey}  \& {Rix}}{{Peng}
  et~al.}{2010}]{2010AJ....139.2097P}
{Peng} C.~Y.,  {Ho} L.~C.,  {Impey} C.~D.,   {Rix} H.-W.,  2010, \mn@doi [\aj]
  {10.1088/0004-6256/139/6/2097}, \href
  {https://ui.adsabs.harvard.edu/abs/2010AJ....139.2097P} {139, 2097}

\bibitem[\protect\citeauthoryear{{Pierre} et~al.,}{{Pierre}
  et~al.}{2004}]{2004JCAP...09..011P}
{Pierre} M.,  et~al., 2004, \mn@doi [\jcap] {10.1088/1475-7516/2004/09/011},
  \href {https://ui.adsabs.harvard.edu/abs/2004JCAP...09..011P} {2004, 011}

\bibitem[\protect\citeauthoryear{Russakovsky et~al.,}{Russakovsky
  et~al.}{2015}]{ILSVRC15}
Russakovsky O.,  et~al., 2015, \mn@doi [International Journal of Computer
  Vision (IJCV)] {10.1007/s11263-015-0816-y}, 115, 211

\bibitem[\protect\citeauthoryear{{Schawinski}, {Zhang}, {Zhang}, {Fowler}  \&
  {Santhanam}}{{Schawinski} et~al.}{2017}]{2017MNRAS.467L.110S}
{Schawinski} K.,  {Zhang} C.,  {Zhang} H.,  {Fowler} L.,   {Santhanam} G.~K.,
  2017, \mn@doi [\mnras] {10.1093/mnrasl/slx008}, \href
  {https://ui.adsabs.harvard.edu/abs/2017MNRAS.467L.110S} {467, L110}

\bibitem[\protect\citeauthoryear{{Shamir}}{{Shamir}}{2017}]{2017PASA...34...44S}
{Shamir} L.,  2017, \mn@doi [\pasa] {10.1017/pasa.2017.40}, \href
  {https://ui.adsabs.harvard.edu/abs/2017PASA...34...44S} {34, e044}

\bibitem[\protect\citeauthoryear{Srivastava, Hinton, Krizhevsky, Sutskever  \&
  Salakhutdinov}{Srivastava et~al.}{2014}]{JMLR:v15:srivastava14a}
Srivastava N.,  Hinton G.,  Krizhevsky A.,  Sutskever I.,   Salakhutdinov R.,
  2014, Journal of Machine Learning Research, 15, 1929

\bibitem[\protect\citeauthoryear{{Strateva} et~al.,}{{Strateva}
  et~al.}{2001}]{2001AJ....122.1861S}
{Strateva} I.,  et~al., 2001, \mn@doi [\aj] {10.1086/323301}, \href
  {https://ui.adsabs.harvard.edu/abs/2001AJ....122.1861S} {122, 1861}

\bibitem[\protect\citeauthoryear{{Sugai} \& {Iye}}{{Sugai} \&
  {Iye}}{1995}]{1995MNRAS.276..327S}
{Sugai} H.,  {Iye} M.,  1995, \mn@doi [\mnras] {10.1093/mnras/276.1.327}, \href
  {https://ui.adsabs.harvard.edu/abs/1995MNRAS.276..327S} {276, 327}

\bibitem[\protect\citeauthoryear{{Tanaka} et~al.,}{{Tanaka}
  et~al.}{2018}]{2018PASJ...70S...9T}
{Tanaka} M.,  et~al., 2018, \mn@doi [\pasj] {10.1093/pasj/psx077}, \href
  {https://ui.adsabs.harvard.edu/abs/2018PASJ...70S...9T} {70, S9}

\bibitem[\protect\citeauthoryear{{White}}{{White}}{1984}]{1984ApJ...286...38W}
{White} S.~D.~M.,  1984, \mn@doi [\apj] {10.1086/162573}, \href
  {https://ui.adsabs.harvard.edu/abs/1984ApJ...286...38W} {286, 38}

\bibitem[\protect\citeauthoryear{{Willett} et~al.,}{{Willett}
  et~al.}{2013}]{2013MNRAS.435.2835W}
{Willett} K.~W.,  et~al., 2013, \mn@doi [\mnras] {10.1093/mnras/stt1458}, \href
  {https://ui.adsabs.harvard.edu/abs/2013MNRAS.435.2835W} {435, 2835}

\bibitem[\protect\citeauthoryear{{York} et~al.,}{{York}
  et~al.}{2000}]{2000AJ....120.1579Y}
{York} D.~G.,  et~al., 2000, \mn@doi [\aj] {10.1086/301513}, \href
  {https://ui.adsabs.harvard.edu/abs/2000AJ....120.1579Y} {120, 1579}

\makeatother
\end{thebibliography}


\bibliographystyle{mnras}




\appendix

\section{Some extra material}\label{sec;SZimages}

\begin{figure*}
\begin{center}
\includegraphics[scale=1.0]{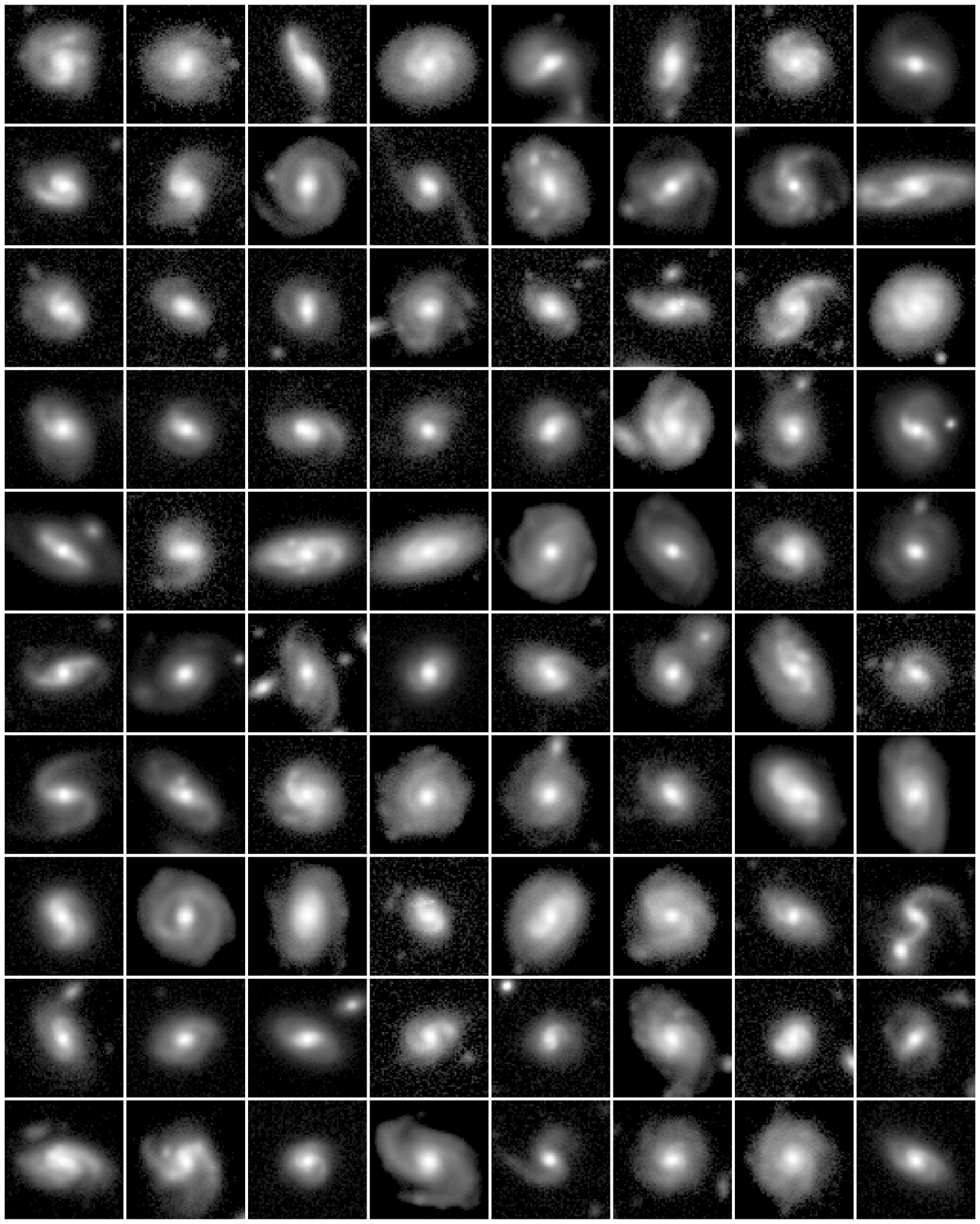}
\end{center}
\caption{Example images of S-spirals.
\label{fig;exam_S}}
\end{figure*}

\begin{figure*}
\begin{center}
\includegraphics[scale=1.0]{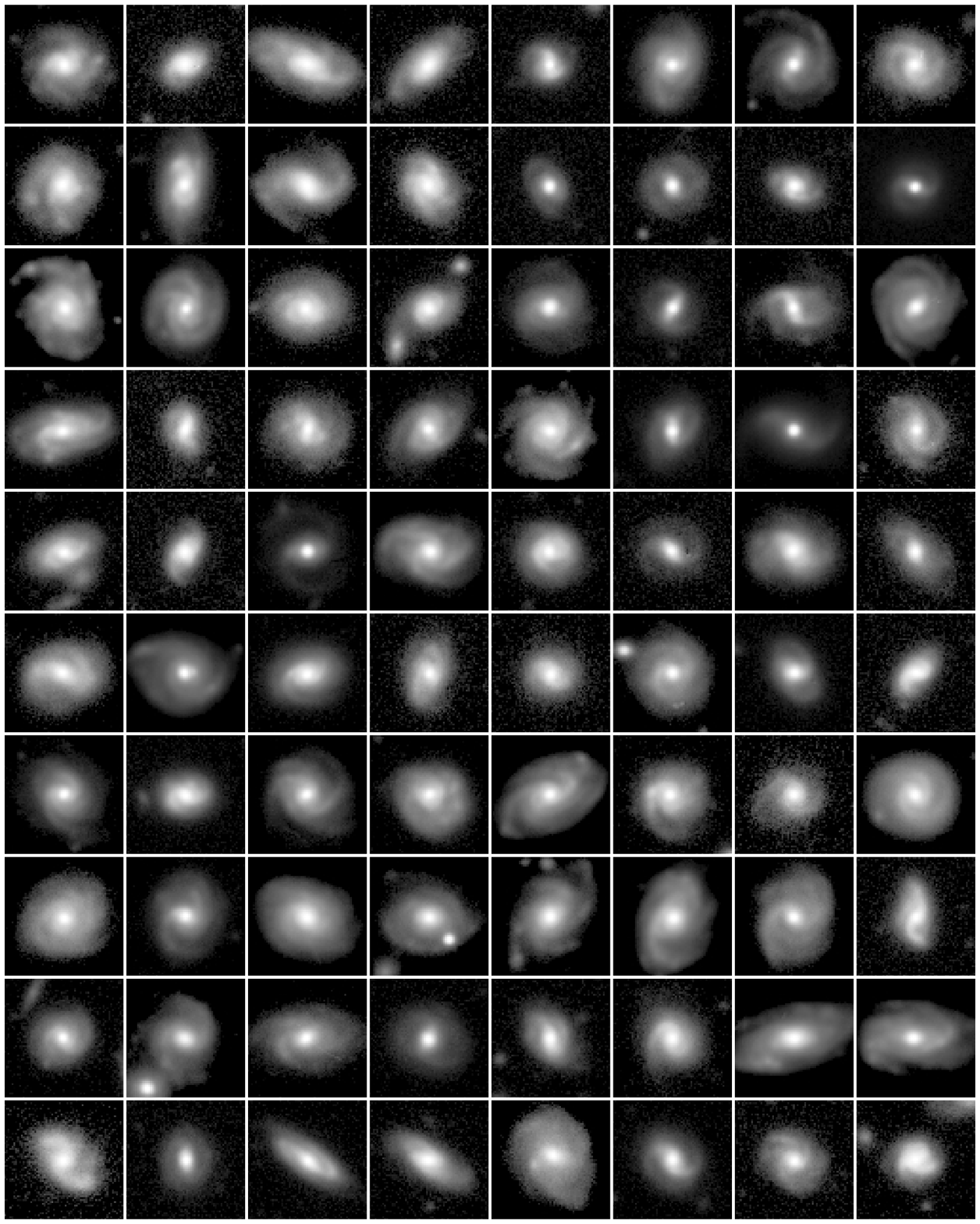}
\end{center}
\caption{Example images of Z-spirals.
\label{fig;exam_Z}}
\end{figure*}

In Figure \ref{fig;exam_S} and Figure \ref{fig;exam_Z}, we show example images of S-spirals and Z-spirals, which are randomly selected from the training data set.
We use them for training a CNN model after data augmentation such as flipping, rescaling, and rotation of images.


\bsp	
\label{lastpage}
\end{document}